\documentstyle[12pt,A4]{article}

\input epsf
\epsfverbosetrue

\begin{document}

\begin{flushright}
HIP -- 1997 -- 55 / TH\\
September 26, 1997
\end{flushright}

\begin{centering}

{\Large\bf A model for multi-quark systems}
\vspace{0.5cm}

A. M. Green\footnote[1]{E-mail: {\tt green@phcu.helsinki.fi}}
and
P. Pennanen\footnote[2]{E-mail: {\tt Petrus@hip.fi}}

\vspace{0.25cm}

{\em Helsinki Institute of Physics and Department of Physics \\
P.O. Box 9, FIN-00014 University of Helsinki, Finland }
\vspace{0.5cm}

{\bf Abstract}

As a step towards understanding multi-quark systems abundant in nature we 
construct a model that reproduces the binding energies of static four-quark 
systems. These energies have been calculated using SU(2) lattice gauge theory 
for a set of six different geometries representative of the general case. The 
model is based on ground and excited state two-body potentials and multi-quark
interaction terms. 


\end{centering}

\noindent

\section{Introduction}

\begin{figure}[htb]
{
\newcommand{\thlen}{\setlength{\unitlength}{0.75pt}}
\newcommand{\quarksTR}{\multiput(0,0)(40,60){2}{\circle*{6}}
	\multiput(60,-20)(40,60){2}{\circle*{6}} }
\newcommand{\quarksNP}{\multiput(0,0)(0,40){2}{\circle*{6}}
	\multiput(60,-20)(40,20){2}{\circle*{6}} }
\newcommand{\quarksT}{\multiput(0,0)(40,60){2}{\circle*{6}}
	\multiput(60,20)(40,-20){2}{\circle*{6}} }
\newcommand{\quarksR}{\multiput(0,0)(0,40){2}{\circle*{5}}
	\multiput(60,0)(0,40){2}{\circle*{5}} }
\newcommand{\cubelabels}[3]{\put(28,-12){\makebox(0,0)[tr]{#1}}
	\put(85,-12){\makebox(0,0)[tl]{#2}}
	\put(105,20){\makebox(0,0)[cl]{#3}} }
\thlen
\newsavebox{\axes}
\savebox{\axes}{ {\thinlines
	\put(0,0){\vector(0,1){65}}
	\put(0,0){\vector(2,1){70}}
	\put(0,0){\vector(3,-1){85}} }}
\newsavebox{\grid}
\savebox{\grid}{ {\thinlines
	\multiput(0,0)(40,20){2}{\line(0,1){40}}
	\multiput(60,-20)(40,20){2}{\line(0,1){40}}
	\multiput(0,0)(60,-20){2}{\line(2,1){40}}
	\multiput(0,40)(60,-20){2}{\line(2,1){40}}
	\multiput(0,0)(0,40){2}{\line(3,-1){60}}
	\multiput(40,20)(0,40){2}{\line(3,-1){60}} }}
\setlength{\unitlength}{0.95pt}
\begin{center}
\setlength{\tabcolsep}{13pt}
\renewcommand{\arraystretch}{3.5}
\begin{tabular}{cccc}
 & \fbox{A} & \fbox{B} & \fbox{C} \\
\raisebox{15pt}{(S,R)} &
\begin{picture}(60,60)
\quarksR
\put(30,48){\makebox(0,0)[b]{$r$}}
\put(22,50){\vector(-1,0){22}}
\put(37,50){\vector(1,0){23}}
\put(67,20){\makebox(0,0)[l]{$d$}}
\put(71,27){\vector(0,1){13}}
\put(71,13){\vector(0,-1){13}}
\put(8,-2){\makebox(0,0)[b]{3}}
\put(52,-2){\makebox(0,0)[b]{4}}
\put(8,35){\makebox(0,0)[b]{1}}
\put(52,35){\makebox(0,0)[b]{2}}
\thicklines
\multiput(0,0)(60,0){2}{\line(0,1){40}}
\end{picture} &
\begin{picture}(60,60)
\quarksR
\put(0,5){\makebox(0,0)[b]{3}}
\put(60,5){\makebox(0,0)[b]{4}}
\put(0,27){\makebox(0,0)[b]{1}}
\put(60,27){\makebox(0,0)[b]{2}}
\thicklines
\multiput(0,0)(0,40){2}{\line(1,0){60}}
\end{picture} &
\begin{picture}(60,60)
\quarksR
\put(0,5){\makebox(0,0)[b]{3}}
\put(60,5){\makebox(0,0)[b]{4}}
\put(0,27){\makebox(0,0)[b]{1}}
\put(60,27){\makebox(0,0)[b]{2}}
\thicklines
\put(0,0){\line(3,2){60}}
\put(0,40){\line(3,-2){60}}
\end{picture} \\

\thlen
\raisebox{40pt}{(TR)} &

\thlen
\begin{picture}(100,95)(0,-30)
\put(0,0){\usebox{\axes}} 
\put(0,0){\usebox{\grid}} 
\quarksTR
\cubelabels{$d$}{$x$}{$y$}
\thicklines
\put(0,0){\line(3,-1){60}}
\put(40,60){\line(3,-1){60}}
\end{picture} &

\thlen
\begin{picture}(100,95)(0,-30)
\put(0,0){\usebox{\axes}} 
\put(0,0){\usebox{\grid}}
\quarksTR
\thicklines
\put(0,0){\line(2,3){40}}
\put(60,-20){\line(2,3){40}}
\end{picture} &

\thlen
\begin{picture}(100,95)(0,-30)
\put(0,0){\usebox{\axes}} 
\put(0,0){\usebox{\grid}} 
\quarksTR
\thicklines
\put(0,0){\line(5,2){100}}
\put(40,60){\line(1,-4){20}}
\end{picture} \\

\raisebox{15pt}{(L)} &
\begin{picture}(80,40)(0,-20)
\multiput(0,0)(30,0){2}{\circle*{5}}
\multiput(50,0)(30,0){2}{\circle*{5}}
\put(15,-4){\makebox(0,0)[t]{$d$}}
\put(65,-4){\makebox(0,0)[t]{$d$}}
\put(25,10){\makebox(0,0)[b]{$r$}}
\put(17,12){\vector(-1,0){17}}
\put(32,12){\vector(1,0){18}}
\thicklines
\multiput(0,0)(50,0){2}{\line(1,0){30}}
\end{picture} &
\begin{picture}(80,40)(0,-20)
\multiput(0,0)(80,0){2}{\circle*{5}}
\multiput(30,-10)(20,0){2}{\circle*{5}}
\thicklines
\put(0,0){\line(1,0){80}}
\put(30,-10){\line(1,0){20}}
\end{picture} &
\begin{picture}(80,40)(0,-20)
\multiput(0,0)(50,0){2}{\circle*{5}}
\multiput(30,-10)(50,0){2}{\circle*{5}}
\thicklines
\multiput(0,0)(30,-10){2}{\line(1,0){50}}
\end{picture} \\

\raisebox{25pt}{(Q)} &
\begin{picture}(60,50)(0,-20)
\multiput(0,0)(30,0){2}{\circle*{5}}
\multiput(60,0)(0,30){2}{\circle*{5}}
\put(15,4){\makebox(0,0)[b]{$d$}}
\put(65,15){\makebox(0,0)[l]{$d$}}
\put(30,-8){\makebox(0,0)[t]{$r$}}
\put(22,-10){\vector(-1,0){22}}
\put(37,-10){\vector(1,0){23}}
\thicklines
\put(0,0){\line(1,0){30}}
\put(60,0){\line(0,1){30}}
\end{picture} &
\begin{picture}(60,40)(0,-20)
\multiput(0,0)(30,0){2}{\circle*{5}}
\multiput(60,0)(0,30){2}{\circle*{5}}
\thicklines
\put(0,0){\line(2,1){60}}
\put(30,0){\line(1,0){30}}
\end{picture} &
\begin{picture}(60,40)(0,-20)
\multiput(0,-5)(60,0){2}{\circle*{5}}
\multiput(30,0)(30,30){2}{\circle*{5}}
\thicklines
\put(0,-5){\line(1,0){60}}
\put(30,0){\line(1,1){30}}
\end{picture} \\

\thlen
\raisebox{40pt}{(NP)} &
\thlen
\begin{picture}(100,95)(0,-30)
\put(0,0){\usebox{\axes}} 
\put(0,0){\usebox{\grid}} 
\quarksNP
\cubelabels{$r$}{$d$}{$d$}
\thicklines
\put(0,0){\line(0,1){40}}
\put(60,-20){\line(2,1){40}}
\end{picture} &

\thlen
\begin{picture}(100,95)(0,-30)
\put(0,0){\usebox{\axes}} 
\put(0,0){\usebox{\grid}}
\quarksNP
\thicklines
\put(0,0){\line(3,-1){60}}
\put(0,40){\line(5,-2){100}}
\end{picture} &

\thlen
\begin{picture}(100,95)(0,-30)
\put(0,0){\usebox{\axes}} 
\put(0,0){\usebox{\grid}} 
\quarksNP
\thicklines
\put(0,0){\line(1,0){100}}
\put(0,40){\line(1,-1){60}}
\end{picture} \\

\thlen
\raisebox{60pt}{(T)} &
\thlen
\begin{picture}(100,110)(0,-55)
\put(0,0){\usebox{\axes}} 
\put(0,0){\usebox{\grid}} 
\quarksT
\put(60,-45){\makebox(0,0)[tr]{\fbox{A}}}
\cubelabels{$r$}{$d$}{$d$}
\thicklines
\put(0,0){\line(3,1){60}}
\put(40,60){\line(1,-1){60}}
\end{picture} &

\thlen
\begin{picture}(100,110)(0,-55)
\put(0,0){\usebox{\axes}} 
\put(0,0){\usebox{\grid}}
\quarksT
\put(60,-45){\makebox(0,0)[tr]{\fbox{B}}}
\thicklines
\put(0,0){\line(1,0){100}}
\put(60,20){\line(-1,2){20}}
\end{picture} &

\thlen
\begin{picture}(100,110)(0,-55)
\put(0,0){\usebox{\axes}} 
\put(0,0){\usebox{\grid}} 
\quarksT
\put(60,-45){\makebox(0,0)[tr]{\fbox{C}}}
\thicklines
\put(0,0){\line(2,3){40}}
\put(100,0){\line(-2,1){40}}
\end{picture}
\end{tabular}
\end{center}

}

\caption{Simulated four-quark geometries and their two-body 
pairings. \label{fconf}}
\end{figure}
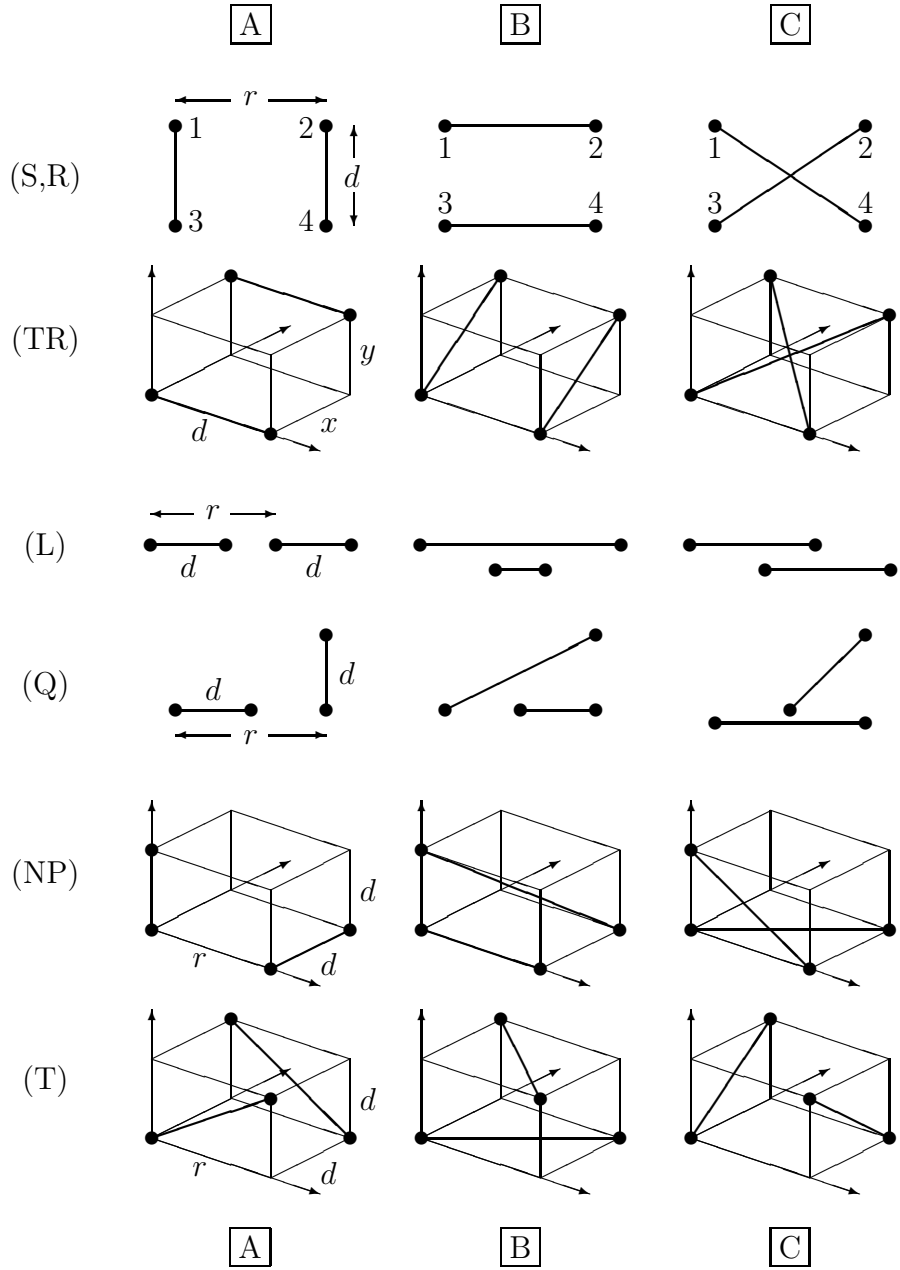

The significant progress in lattice QCD has so far been restricted to systems  with only a few quarks -- up to three
in most cases. However, in particle and nuclear
physics there is also considerable interest in few- and multi-{\em hadron}
systems beginning with the possibility of bound $K\bar{K}$ states. 
This prompts one to ask what lattice QCD can say about these more
complex quark systems. It seems unlikely that, in the foreseeable
future, the techniques of lattice QCD can be developed
sufficiently to tackle these problems directly. Therefore, we are reduced
to constructing models that can explain the lattice results for the simplest
multi-quark systems in a way that can be readily extended to more complicated
cases. With this in mind, the Helsinki group in 
Refs. \cite{gmp:93,gmps:93,gms:94}
has calculated the energies of four quarks on a lattice in various
geometries -- namely, four quarks at the corners of rectangles,
tetrahedra and other geometries as shown in Fig. \ref{fconf}. This is 
taken to be our 'experimental' data, which is to be explained by some model.
Hopefully, this set of geometries is general enough for
the model to also explain the energies of geometries
in between those actually considered. Our philosophy 
is that, if {\em any} geometry cannot be fitted, then the model fails, 
since then there is no reason to  expect configurations not checked 
explicitly to be fitted.

Due to limitations of computing resources the 'experimental' data is not for 
full QCD. Instead of SU(3) we use SU(2), which saves about an order of 
magnitude in computer time. The less than 10\% quantitative differences
observed in the glueball spectrum and topological susceptibility of these 
theories suggest that the 
relevant qualitative features are preserved \cite{tep:97}. Secondly, since the
quarks are fixed in some geometry, they are static. Therefore this
discussion is most relevant to the heaviest quarks, in analogy to the static
potentials used successfully to describe quarkonia. The
static approximation is currently being partially removed by 
applying the  techniques of 
Ref. \cite{mic:97} to a system of two B mesons with only the $b$ quarks 
static. 
A third approximation is that no quark-pair creation is allowed the so-called
quenched approximation. The effects from this approximation have been 
found to be of the same magnitude as the use of SU(2) instead of SU(3), apart 
from string breaking effects at distances larger than we simulate. 
We concentrate here on 
attempting to understand the best simulation data available, i.e. for a system 
with four static SU(2) quarks in the 
quenched approximation. 

Any model that can be extended to multi-quark systems must presumably
treat only the quark degrees of freedom explicitly -- with the gluon
degrees of freedom entering only implicitly. This is the same philosophy
as used, with much success, for interacting multi-nucleon systems. In the
latter, the meson fields generating the interactions lead to effective
internucleon potentials, which take the form of two-nucleon potentials
and, to a lesser extent, three- and four-nucleon potentials. However, it
must be remembered that multi-quark and multi-nucleon systems are very
different, since in the former the underlying gluon fields are 
strongly self-interacting. It is, therefore, not at all clear that {\em
any} effective model defined in terms of only quark degrees of
freedom will be successful. It is the purpose of this article to see to
what extent such a model can be developed  for the four-quark case. If
this trial fails, then there is no point in expecting it to work in even
more complex quark problems i.e. a successful model for four quarks is
necessary but not sufficient before considering any extension to even
more quarks.

In Section 2 the model is introduced. The results are given in Section 3 and
concluding remarks presented in Section 4.

\section{The model}

The basis $A,B,\ldots$ of the model contains pairings of the quarks in 
all possible ways. A normalization matrix ${\bf N}$ contains the overlaps
of any two states, $N_{AA} = <A|A>,\; N_{AB} = <A|B>, \ldots$, while a 
potential matrix ${\bf V}$ for an interaction $V$ is defined as 
$V_{AA} = <A|V|A>,\; V_{AB} = <A|V|B>, \ldots$. The object is then to 
compare the eigenvalues
\begin{equation}
\label{VN}
\left[{\bf V}-E(4) {\bf N}\right]=0
\end{equation}  
 with the lattice
results -- the success or failure of the model being to what extent the
two can be made to agree. 

As we only consider SU(2),
there is no distinction between the group properties of quarks ($q$) and
antiquarks ($\bar{q}$).  As shown in Fig. \ref{fconf}, four such quarks can 
then be paired in three different ways
\begin{equation}
\label{ABC}
A=(q_1q_3)(q_2q_4), \ \ B=(q_1q_2)(q_3q_4) \ \ {\rm and} \ \ C=(q_1q_4)(q_2q_3),
\end{equation}
where each $(q_iq_j)$ is a colour singlet. These three basis
states are not orthogonal to each other and have in the weak coupling limit
the condition \cite{gre:88}
\begin{equation}
\label{A+B+C}
|A+B+C>=0.
\end{equation}
Since $<A|A>=<B|B>=<C|C>=1$, we get in this limit the equalities  \\
$<A|B>=<B|C>=<A|C>=-1/2$.

The form of the potential is motivated by perturbative one-gluon exchange
i.e.
\begin{equation}
\label{OGE}
V_{ij}=-\frac{1}{3}\sum_{i\le j} \tau_i \tau_j v_{ij},
\end{equation}
where, for the ground state, we use the Cornell form of parameterization of 
the static two-quark potential
\begin{equation}
v_{ij}=-e/r_{ij}+b_sr_{ij}+c \label{V2}
\end{equation}
fitted to measured lattice values.
Then $V_{AA}=v_{13}+v_{24}$ etc. and
\begin{equation}
\label{VAB}
V_{AB}=V_{BA}= <A|V|B> = 
-\frac{1}{2}\left(v_{13} +v_{24} +v_{12}+v_{34} - v_{14}-v_{23} \right).
\end{equation}
Here it should be noted that using this perturbative form of the potential
without a multi-quark interaction term is unacceptable because of the resulting
unphysical long range Van der Waals forces. When moving to stronger couplings
the overlap between any two states should decrease, becoming zero in the 
strong coupling (or large distance) limit. We take this effect into account
by introducing a factor $f$, defined as $<A|B>=-f/2$, which decreases with
increasing separation of the quarks.

For internal consistency, this same factor must also multiply
$V_{AB}$ in Eq.~\ref{VAB} -- otherwise the eigenvalues would depend on the 
self-energy term $c$ in Eq.~\ref{V2}. A reasonable parameterization for the 
multi-quark interaction term $f$ is
\begin{equation}
\label{f}
f=\exp(-b_s k_f S),
\end{equation}
where $b_s$ is the string energy of Eq.~\ref{V2}, $S$ the minimum area bounded
by the straight lines connecting the quarks and $k_f$ a parameter to be fitted.

This form of the parameterization was originally 
motivated by strong coupling ideas \cite{mor:89,ale:90}. More complicated
parameterizations, such as 
$f=f_0 \exp(-b_s k_f S-\sqrt{b_s} k_P P)$ with $P$ the perimeter of $S$ and
$f_0,k_P$ new constants to be fitted, have been suggested. However, the new
parameters were found to measure lattice artefacts in a continuum
extrapolation, where $f_0\rightarrow 1,\; k_P \rightarrow 0$ \cite{pen:96b}.
The determination of a minimal area $S$ is a laborious task \cite{fur:96} and 
thus impractical when many such areas are needed. Therefore, we use a 
formula that is readily evaluated for any geometry; the area is simply taken
to be the average of the sum of the four triangular areas defined by  the
positions of the four quarks i.e. the faces of the tetrahedron. For
example, in the notation of Eq.~\ref{ABC}, the appropriate area $S(AB)$ for $f$ is
\begin{equation}
\label{Def:S}
S(AB)= 0.5[S(431)+S(432)+S(123)+S(124)],
\end{equation}
where $S(ijk)$ is the area of the triangle with corners at $i,j$ and
$k$. For planar geometries this simply reduces to the expected area, while
for non-planar cases this is only an approximation to $S(AB)$.

Of the six geometries considered here the tetrahedron presents the most 
challenging data. Other geometries could be fitted qualitatively with the above
model using only the one free parameter $k_f$ in Eq.~\ref{f}. However,
the regular tetrahedron has maximal degeneracy as all quarks are at an equal
distance from each other. Its energies have the interesting feature
that the lowest state is doubly degenerate and becomes {\em more} bound as 
the tetrahedron
increases in size, as noted in Ref. \cite{glpm:96}. This is opposite
to what happens with squares, where the magnitude of $E_0$ decreases with
increasing size of the system. This indicates that there is
coupling to some higher state(s) that becomes more effective as the size
increases and suggests that these higher states contain gluon
excitation with respect to the basic $A,B,C$ configurations. Therefore, as 
proposed in Ref. \cite{pen:96b}, we include
states, denoted with a prime, which describe the same quark partitions but
with the gluonic field in an excited state. These excited states could be
excitations of the two-body potential or flux configurations where the four 
quarks form a singlet. In the latter case the formalism below needs to be 
slightly modified. If the excited state is taken as the lowest two-body 
excitation, for which the gluonic
field has the symmetry of the $E_u$ representation of the lattice symmetry
group $D_{4h}$, we have
\begin{equation}
A^*=(q_1 q_3)_{E_u} (q_2 q_4)_{E_u} \ \ {\rm etc}.
\end{equation}
Because the $E_u$ state is an odd parity excitation,
$A^*,B^*,C^*$ must contain two such states in order to have the
same parity as $A,B,C$.
The excitation energy of an $E_u$ state over its ground state($A_{1g}$)
counterpart is $\approx \pi/R$ for two quarks a distance $R$ apart.
As said above, when $R$ increases this excitation energy decreases making
the effect of the $A^*,B^*,C^*$ states more important. 

In analogy to Eq.~\ref{A+B+C} an antisymmetry condition $A^*+B^*+C^*=0$ holds
for the excited states. As the gluonic excitations are orthogonal to 
the the ground state we also have $<A|A^*>=<B|B^*>=<C|C^*>=0$. There are now 
two more functions $f^{a,c}$ defined as
\[<A^*|B^*>=<A^*|C^*>=<B^*|C^*>=-f^c/2  \ \ {\rm and} \ \ \]
\begin{equation}
<A^*|B>=<A^*|C>=\ldots \ \ {\rm etc} \ \ \ldots=-f^a/2.
\end{equation}
Here it is assumed that $f^{a,c}$ are both dependent on $S$ as defined in
Eq.~\ref{Def:S}. Since $f^c$  involves only the excited states, it is
reasonable to expect it has a form similar to $f$ in Eq.~\ref{f} i.e.
\begin{equation}
\label{fc}
f^c=\exp(-b_sk_cS).
\end{equation}

In the weak coupling limit, from the $A^*+B^*+C^*=0$ condition,
we expect $<A|B^*>=<B|C^*>=.....=0$ at small distances. This also happens
in the Isgur-Paton model, as can be seen in Fig. 6 of Ref. \cite{pen:96b}.
To take this into account we parameterize $f^a$ as
\begin{equation}
\label{fa}
f^a=(f^a_0+f^a_S S)\exp(-b_s k_a S). 
\end{equation}
Justification of this is given by the Isgur-Paton calculation and an analysis
of the correlations between the paths involved in the simulation, which gives 
the term linear in $S$. When all three parameters are varied in a 
{\tt Minuit} fit it is found that 
$f^a_0=0.002(30)$ is consistent with zero as expected. Therefore, from now on 
we set this parameter to zero.

For the potential matrix ${\bf V(f)}$ the diagonal matrix elements, 
before the lowest energy amongst the
basis states -- usually $V_{AA}$ -- is removed, are
\[<A^*|V|A^*>=v^*(13)+v^*(24), \ \ {\rm etc},\]
where  $v^*(ij)$ is the potential of the $E_u$ state -- a quantity also
measured on the lattice along with the four-quark energies. However, to
allow more freedom in the following fits, we introduce a parameter $b_0$
in
\begin{equation}
<A^*|V|A^*>= V_{AA}+b_0 V^*_{AA}; \ \ 
V^*_{AA}=v^*(13)+v^*(24)-v(13)-v(24) \label{C*C*}
\end{equation}
for state $A^*$ and analogously for $B^*,C^*$. 

If $E_u$ is the most important excitation in the parameter fit, $b_0$ should 
turn out to be of the order unity.
However, in practice we get 2.3(5), suggesting that excitations with higher
energies are, perhaps, more relevant.
 
The two-quark potentials $v(ij)$ are taken to be more elaborate than the
three term form of Eq.~\ref{V2}. They are fitted to the lattice data
using \cite{gre:96}
\begin{equation}
\label{Vfit}
v(r_{ij})=0.562 + 0.0696 r_{ij}-  \frac{0.255}{r_{ij}} -
\frac{0.045}{r^2_{ij}}.
\end{equation}
Similarly, the excitation of the $E_u$ state is fitted by 
\begin{equation}
\label{V*fit}
v^*(ij)-v(ij) =
\frac{\pi}{r_{ij}} - \frac{4.24}{r^2_{ij}} + \frac{3.983}{r^4_{ij}}.
\end{equation}
The extra terms containing $r^{-2}_{ij}$ and $r^{-4}_{ij}$ are purely for numerical
reasons and ensure that the fitted values of $v(ij)$ and $v^*(ij)$ are,
on average, well within 1\% of the lattice values for all $r_{ij}\ge 2$.

There are two types of off-diagonal element
\begin{equation}
\label{off}
<A^*|V|B^*> \ \ {\rm and} \ \ <A|V|B^*>.
\end{equation}
These require further approximations. Using the Isgur-Paton model with N=1
as a guide we get {\em qualitatively}
\begin{equation}
<A^*|V|B^*>=-\frac{ f^c}{2}\left[V_{AA}+V_{BB}-V_{CC}+
c_0\frac{(V^*_{AA}+V^*_{BB})}{2}\right]=<B^*|V|A^*>
\end{equation}
and analogously for the elements $<A^*|V|C^*>, <B^*|V|C^*>$.

Likewise,
\[<A|V|B^*>=-\frac{f^a}{2}\left[V_{AA}+V_{BB}-V_{CC}+a_0\frac{V^*_{BB}}{2}
\right]\] 
\begin{equation}
\label{A*B}
<A^*|V|B>=-\frac{f^a}{2}\left[V_{AA}+V_{BB}-V_{CC}+a_0\frac{V^*_{AA}}{2}
\right] \ \ {\rm etc,}
\end{equation} 
where $a_0,c_0$ are free parameters, which should have values of order
unity if $E_u$ is the most relevant excitation.

In the special case of regular tetrahedra,  ${\bf V}$ reduces to the form

\begin{equation}
\label{KH2}
{\bf V}= \left[ \begin{array}{ccc|ccc}
V_{AA}&-fV_{AA}/2&-fV_{AA}/2&0&-f^aV_a/2&-f^aV_a/2 \\
-fV_{AA}/2&V_{AA}&-fV_{AA}/2&-f^aV_a/2 &0      &-f^aV_a/2     \\
-fV_{AA}/2&-fV_{AA}/2&V_{AA}&-f^aV_a/2 & -f^aV_a/2     &0          \\  \hline
0 &  -f^aV_a/2  & -f^aV_a/2   &V_b&-f^cV_c/2&-f^cV_c/2          \\
-f^aV_a/2 &0    & -f^aV_a/2   &-f^cV_c/2 &V_b     &-f^cV_c/2             \\
-f^aV_a/2 & -f^aV_a/2   &0    &-f^cV_c/2 &   -f^cV_c/2   &V_b       \\ 
\end{array} \right],
\end{equation}

where $V_a=V_{AA}+a_0V^*_{AA}/2 \ , \ V_b=V_{AA}+b_0V^*_{AA} \ , \
V_c=V_{AA}+c_0V^*_{AA}.$ As with all geometries 

\begin{equation}
\label{KH1}
{\bf N}= \left[ \begin{array}{ccc|ccc}
1&-f/2  &-f/2  &0&-f^a/2&-f^a/2 \\
-f/2&1 &-f/2  &-f^a/2 &0     &-f^a/2    \\
-f/2 & -f/2  &1  &-f^a/2 &-f^a/2      &0        \\     \hline
0&-f^a/2&-f^a/2&1&-f^c/2&-f^c/2         \\
-f^a/2 &0     &-f^a/2&-f^c/2 &1     &-f^c/2            \\
-f^a/2 &-f^a/2      &0     &-f^c/2 & -f^c/2     &1         \\      
\end{array} \right].
\end{equation}

The full $6\times 6$ matrix $[{\bf V}-E{\bf N}]$ now breaks into three
$2\times 2$ matrices,
two of which are identical -- giving the observed degeneracy.
These have the form
\begin{eqnarray}
[{\bf V} & - & E{\bf N}] = 
  \left[ \begin{array}{cc}
-E(1+f/2)&-f^a(E-V_a)/2    \\
 -f^a(E-V_a)/2  &-E(1+f^c/2)+V_b+f^cV_c/2 \\
\end{array} \right]=0 
\end{eqnarray}
whereas the third $2\times 2$ matrix giving nondegenerate energies is

\begin{equation}
[{\bf V}-E{\bf N}]= \left[ \begin{array}{cc}
-E(1-f)&f^a(E-V_a)    \\
 f^a(E-V_a)  &-E(1-f^c)+V_b-f^cV_c \\
\end{array} \right]=0. 
\end{equation}

\section{Results}

In Refs. \cite{gmp:93,gmps:93,gms:94,glpm:96} four quark energies have been 
extracted for a variety of geometries using a $16^3\times 32$ lattice with 
$\beta=2.4$. This $\beta$ value corresponds to a lattice spacing $a=0.119(1)$ 
fm. From these energies, one 
hundred -- distributed over all measured geometries -- are selected for 
fitting. Configurations containing flux links of less than two lattice
units were not included because of the strong lattice artefacts 
they contain. The
six geometries are shown in Fig. 1. Specifically, we use 15 Tetrahedra (T), 6
Squares (S), 12 Rectangles (R) (including Tilted Rectangles (TR)), 4 
Quadrilaterals (Q), 9 Non-Planar (NP) and 4 Linear (L).
Only the lowest two energies $(E_{0,1})$ from the lattice simulation are used.
In most cases a three basis simulation had been performed, so that a third
energy $(E_2)$ was in fact available. However, as this state is the
highest calculated, it is not expected to be very reliable due to the 
higher excitations it contains. Its main purpose
was to improve the estimate on $E_1$ by reducing its excited state 
contamination. One might question the connection of these energies with the 
continuum values. This has been answered in Ref. \cite{pen:96b}, 
where it was found that the binding energies for equal physical sizes 
essentially stay 
constant when the lattice spacing is made smaller.

Before commencing a fit, the size of the errors on the above data
must be decided. The lattice simulation, through the boot-strap method,
does indeed produce errors -- statistical ones. However, some estimate
must also be added for systematic errors. How this is done is somewhat
subjective. Here the prescription is to assume all errors must be at
least 0.005 and, also, at least 10\% of the eigenvalue itself. 
The former corresponds to about 10\%, 1\% for the largest values of 
$E_0, E_1$ respectively.

The above 100 pieces of data were fitted with {\tt Minuit} -- the Migrad option
-- using the seven parameters: $k_f$ in  Eq.~\ref{f}, $f^a_S,k_a$ in 
Eq.~\ref{fa}, $k_c$ in  Eq.~\ref{fc}, $b_0$ in Eq.~\ref{C*C*} and $a_0, c_0$ 
in Eqs.~\ref{A*B}. The outcome yielded a 
$\chi^2/$d.o.f.=1.08 with the values of the parameters being given in
 Table~\ref{minuit}.
\begin{table}[htb]
\begin{center}
\begin{tabular}{l|ccccccc}
Parameter & $k_f$ & $k_a$  & $f^a_S$& $k_c$  &$a_0$ &$b_0$ &$c_0$ \\ \hline
Value     &1.25(6)&0.54(11)&0.046(3)&0.04(20)&4.4(3)&2.2(6)&8.0(4) \\ 
\end{tabular}
\caption{The values of the parameters defining the interaction}
\label{minuit}
\end{center}
\end{table}

Of these values:

a) The observation that $k_c\approx 0$ implies that $f^c\approx 1$ i.e. the
excited configurations interact amongst themselves in the way expected from
perturbation theory.

b) The values of $b_0$ (along with $a_0, c_0$) are somewhat larger than the 
naively expected value of unity. This suggests that higher order 
effects are important.

Table ~\ref{chi} shows  the contributions to the
total $\chi^2$/d.o.f. from
each of the 12 types of data -- i.e. from $E_0,E_1$ for the six
geometries ($T,S,R,Q,NP,L$). For comparison the $2\times 2$ basis model ($A,B$)
with $f=1$ is also shown. This corresponds to the absence of multi-quark 
interaction. 
In spite of the frequent use in the literature of such models based only on 
two-body potentials this seems to be a very poor choice.

\begin{table}[htb]
\begin{center}
\begin{tabular}{l|cc|cc}
Basis   & \multicolumn{2}{c|}{$6\times 6$} & \multicolumn{2}{c}{$2\times 2$} \\ \hline
Geometry& $\chi^2(E_0)$&$\chi^2(E_1)$ & $\chi^2(E_0)$&$\chi^2(E_1)$\\ \hline 
T&0.18&0.20&2.4&21.4\\
S&0.01&0.11&6.2&40.0\\
R&0.17&0.09&9.6&63.9\\
Q&0.09&0.03&0.03&0.25\\
NP&0.08&0.01&0.06&1.53\\
L&0.01&0.11&0.01&0.10\\ 
\end{tabular}
\caption{The contributions of the two states ($E_0,E_1$) of each
of the six geometries to the total $\chi^2$/d.o.f. of 1.08 
($6\times 6$ basis) and 146 ($2\times 2$). }

\label{chi}
\end{center}
\end{table}
 
\section{Discussion}

The above model with 6 basis states fits well all simulated geometries and 
confirms our earlier work that a $2\times 2$ basis model with only two-quark 
interactions (i.e. $f=1$) is not able to even qualitatively account for the 
data. Areas of further study include
the actual nature of the higher excitations, which seem to be playing an important 
role. A natural 
extension of the present static quark model is the application of the model 
in a dynamic case using 
the Schr\"odinger equation. Work on simulating such a system is in 
progress.

\section{Acknowledgement}

We thank J. Lukkarinen for sharing his \LaTeX~art. 
Funding from the Finnish Academy and M. Ehrnrooth foundation (P.P) is 
gratefully acknowledged. Our simulations were performed at the Center 
for Scientific Computing in Espoo, Finland.

\end{document}